\documentclass[twocolumn]{elsart}

\usepackage{natbib}

\usepackage{amssymb}

\begin{document}

\begin{frontmatter}


\title{Modeling the X-ray contribution of XRB jets} 
\author[MIT]{Sera Markoff\corauthref{cor1}}, 
\corauth[cor1]{NSF Postdoctoral Fellow, corresponding author.  Email:
sera@space.mit.edu}
\author[MIT]{Michael Nowak},
\author[Saclay]{St\'ephane Corbel},
\author[Amst]{Rob Fender} \&
\author[MPI]{Heino Falcke}
\address[MIT]{MIT, Center for Space
Research Rm. NE80-6035, USA}
\address[Saclay]{Universit\'e Paris VII and Service d'Astrophysique,
CEA, France}
\address[Amst]{University of Amsterdam, The Netherlands}
\address[MPI]{Max-Planck-Institut f\"ur Radioastronomie, Germany}

\begin{abstract}
Astrophysical jets exist in both XRBs and AGN, and seem to share
common features, particularly in the radio.  While AGN jets are known
to emit X-rays, the situation for XRB jets is not so clear.  Radio
jets have been resolved in several XRBs in the low/hard state,
establishing that that some form of outflow is routinely present in
this state.  Interestingly, the flat-to-inverted radio synchrotron
emission associated with these outflows strongly correlates with the
X-ray emission in several sources, suggesting that the jet plasma
plays a role at higher frequencies.  In this same state, there is
increasing evidence for a turnover in the IR/optical where the
flat-to-inverted spectrum seems to connect to an optically thin
component extending into the X-rays.  We discuss how jet synchrotron
emission is likely to contribute to the X-rays, in addition to inverse
Compton up-scattering, providing a natural explanation for these
correlations and the turnover in the IR/optical band.  We present
model parameters for fits to several sources, and address some common
misconceptions about the jet model.
\end{abstract}

%
%

\end{frontmatter}

\vspace{-.3in}
\section{Introduction}
\label{intro}

\vspace{-.3in}
While the jets in Active Galactic Nuclei (AGN) are known to radiate
from the radio through the X-rays, the jets associated with low/hard
state (LHS) X-ray binaries (XRBs) have only been conclusively
identified as the source of the radio emission for a few sources
\citep{Fender2001}.  However, these smaller jets do share the
``signature'' flat-to-inverted radio synchrotron spectrum seen from
the compact inner jets in AGN, so it is likely that the same emission
mechanism is at work even when the jets are as yet unresolved.  Unlike
in AGN, however, this optically thick synchrotron can extend well into
the infrared (IR) and beyond \citep[e.g.,][]{CorbelFender2002}.

\vspace{-.1in}At higher frequencies, the picture is not so clear.  Although AGN jets
contribute to the X-rays via both synchrotron and inverse Compton (IC)
emission, these processes have not been considered in the standard
picture for the high-frequency emission of XRBs.  Beyond the IR, a
typical LHS spectrum shows a weak thermal contribution plus a hard
power-law extending to typically $\sim100$ keV.  These have generally
been interpreted in terms of a Standard Thin Disk
\citep[SD;][]{ShakuraSunyaev1973} in combination with an optically
thin flow at smaller radii, or an overlying corona
\citep[for reviews see][]{Poutanen1998, NowakWilmsDove2002}.  In this
scenario, the hard power-law originates in IC upscattering of the
thermal SD photons by the hotter plasma.

\vspace{-.1in}Although variations of this picture can successfully explain the X-ray
features, they cannot address the very tight correlation between the
X-ray and radio in the LHS extending down to the Quiescent State (QS),
which has now been seen in several sources
\citep[e.g.][]{Corbeletal2002, GalloFenderPooley2002}. Similarly, in
all XRBs with quasi-simultaneous or better broadband data, there seems
to be a ``turnover coincidence'' in which the X-rays always trace back
to the same decade of frequency between $\sim10^{14}-10^{15}$ Hz in
the IR.  This suggests some connection between the low and high
frequencies which is present for all LHS sources.

\vspace{-.1in}An alternative model for the LHS emission, involving a combination of
jet and thermal SD emission, has been applied to several sources
\citep[e.g.,][]{MarkoffFalckeFender,Markoffetal2003}.  It is currently
the only model which can simultaneously fit the broadband spectrum,
account for the radio/X-ray correlations (as well as analytically
explain the slope of the correlations) and explain the turnover
coincidence between the IR and X-rays.  The X-rays can stem from
synchrotron, IC, or both, processes depending on maximum energy to
which the particles can be accelerated in the jet, which itself
depends on cooling processes.

\vspace{-.1in}
The details of the jet model have been discussed elsewhere, so we
would like to briefly touch on the main points as well as a few common
misconceptions about the workings of the model.

\vspace{-.3in}\section{General results \& discussion}

{\small 
\begin{table*}[t]
\caption{Jet model parameters for several sources.  The mass, distance
and inclination are taken from observations, when available.  The
input total power in the jet, the initial electron plasma temperature
and the location of the acceleration zone are found from fitting.}
\begin{tabular}{|l|c|c|c|c|c|c|}
\hline
{\bf Source} & {\bf M$_{\rm bh}$ (M$_{\odot}$)} & {\bf D (kpc)} & {\bf $\theta_i^\circ$} &
{\bf Jet power (L$_{\rm Edd}$)} & {\bf T$_e$ (K)} & {\bf z$_{\rm sh}$ (r$_g$)}\\
\hline
\vspace{-.1in}XTE~J1118+480 & 6 & 1.8 & 31 & $1.5\times10^{-2}$ & $7\times10^9$ &
150 \\
\vspace{-.1in}GX~339$-$4 (1981) & 5 & 4.0 & 55 & $2.0\times10^{-1}$ & $7\times10^9$ &
$1750$\\
\vspace{-.1in}Cyg X-1$^*$ & 10 & 2.5 & 35 & $1.4\times10^{-2}$& $7\times10^9$ & $47$\\
\vspace{-.1in}V404~Cyg$^{**}$ & 12 & 3.5 & 56 & $1.4\times10^{-1}$ & $7\times10^9$ &
670\\
\vspace{-.1in}GRO~J0422+42 & 5 & 2.5 & 41 & $1.5\times10^{-2}$ & $2\times10^{10}$ &
100\\
\vspace{-.1in}XTE J1550-564 & 10 & 4.0 & 75 & $3.5\times10^{-2}$ & $7\times10^9$ &
215\\
XTE J1650-500 & 10? & 3.0? & 57? & $3.0\times10^{-2}$ &
$2.0\times10^{10}$ & $9000$ \\
\hline
\end{tabular}

{\small $^*$ Non-simultaneous data set, $^{**}$ No spectral info for X-ray flux}
\end{table*}
}

\vspace{-.2in}We present the important model parameters for several sources in
Tab. 1.  While these results are preliminary in that they do not yet
address the fine features in the X-rays (e.g., line emission and
reflection components), some basic trends are revealing themselves.
The jets can account for the entire broadband spectrum with 1-10\% of
the total Eddington luminosity of the source, assuming relativistic
thermal electrons in the accreting plasma, and that a fraction of this
plasma is accelerated.  The location where the acceleration starts
falls between $\sim 10^2-10^3$ $r_g$ with the exception of two
sources.  The first is Cyg X-1, where the data are not simultaneous,
and for which we are currently working with a newer data set
(R. Fender, priv.comm.) which will give better restraints.  The binary
parameters of the other outlier, XTE J1650-500, are still
unconstrained so we assumed reasonable values, which however could be
contributing to this difference.

\vspace{-.1in}The location of $z_{\rm sh}$ is determined by fitting the turnover
from the optically thick IR to the optically thin X-ray regime in the
jet.  The range of fit values is very small, considering that the jet
must extend at least out to $\sim10^{10}$ $r_g$ to fit the radio.
This reflects the ``turnover coincidence'' seen in all LHS sources
with flat-to-inverted radio and quasi-simultaneous X-rays, and
suggests that all sources share a common structure where acceleration
occurs.  Synchrotron from the jet is a natural explanation for this
coincidence, and can similarly explain the radio/soft X-ray
correlations.  If the synchrotron only extends into the soft X-rays
and IC from the base of the jet dominates at hard X-rays, the model
could also address the turnover coincidence and correlations.
Interestingly, this turnover was explicitly resolved in GX~339$-$4
\citep{CorbelFender2002}.  Furthermore, fits to spectral data from
massive jets in blazars also show a similar scale for the dominant
emission region \citep{Beckmannetal2002}, strengthening the argument
that the basic physics of jets scale to some extent.

\vspace{-.1in}What is becoming clear is that the processes behind the radio/IR and
X-ray emission can no longer be considered independent.  This is not,
however, incompatible with the standard picture if one considers that
the idea of a hot, magnetized corona above the SD could be synonymous
or the source of the jet base \citep[see, e.g.,][]{MerloniFabian2002} .  Unifying these two types of models
provides a challenge, however.  In the case of GX~339$-$4, for
example, a sphere-disk model can well explain the X-ray spectrum in
many details \citep{NowakWilmsDove2002}, but not the radio/IR and the
correlations.  If the jet can explain the entire broadband spectrum,
and if its base is somehow equivalent to, or contiguous with, the
corona, there must be a self-consistent parameter range for combining
these two pictures.  This is problematic at first consideration,
however, since the temperatures required at the jet model base are
much higher than ``typical'' coronal values, while the scale of the
jet base is a factor of $\sim10$ smaller.  It is possible that by
combining these two scenarios, both models will be required to explore
new parameter ranges to find a compatible solution.  This is work we
are currently exploring.

\vspace{-.3in}\section{Addressing some common misconceptions}

\vspace{-.2in}As a final note, we have noticed several misconceptions about the
nature of the jet model in the literature and in the remaining space
would like to touch on a few of these.  

\vspace{-.1in}First, spectral pivoting in the X-ray would not result in orders of
magnitude variability in the radio, which is of course not observed.
If one takes self-absorption into account, as is integral to our
model, the X-ray pivoting requires mainly a change in the location of
the optically thick-to-thin turnover.  Another point raised is that
the jet model is incompatible with reflection features.  As seen in
Tab. 1, the X-ray emission originates close enough to the disk that
beaming away from the disk is weak because near the base of the jet
the plasma is near the local speed of sound.  It is also important to
note the increasing evidence for misaligned jets in XRBs, where the
jet is in fact inclined towards the disk.  This, along with disk
warping, means that jets may be able to reproduce most reflection
features.  Along these lines, the location of the X-ray emitting
region is also compatible with estimates based on the amplitude of
reflection and the iron line.  Finally, it has been argued that the
shape of the cutoff itself argues against a non-thermal
interpretation.  The jet model can naturally explain the cutoff via
synchrotron emission or IC from the jet---which process dominates is
dependent on the photon field available for upscattering.  Arguments
based strictly on the shape of the cutoff are not necessarily valid,
because the cutoff data is likely averaged over many variations
\citep[e.g.,][]{PoutanenFabian1999}, and the statistics are not as
good.


\end{document}